\documentclass[conference]{IEEEtran}
\IEEEoverridecommandlockouts
\usepackage{cite}
\usepackage{amsmath,amssymb,amsfonts}
\usepackage{algorithm}
\usepackage{algpseudocode}
\usepackage{mathtools}
\usepackage{tikz,pgfplots}
\usepackage{graphicx}
\usepackage{textcomp}
\usepackage{xcolor}
\usepackage{hyperref}

\def\BibTeX{{\rm B\kern-.05em{\sc i\kern-.025em b}\kern-.08em
    T\kern-.1667em\lower.7ex\hbox{E}\kern-.125emX}}

\DeclarePairedDelimiter\floor{\lfloor}{\rfloor}

\begin{document}

\title{Multithreaded Filtering Preconditioner for Diffusion Equation on Structured Grid}


\author{\IEEEauthorblockN{Abhinav Aggarwal\IEEEauthorrefmark{1},
Shivam Kakkar\IEEEauthorrefmark{2}, Pawan Kumar\IEEEauthorrefmark{3}}
\IEEEauthorblockA{International Institute Of Information Technology, Hyderabad, India \\
Email: \IEEEauthorrefmark{1}abhinav.aggarwal@students.iiit.ac.in,
\IEEEauthorrefmark{2}shivam.kakkar@students.iiit.ac.in,
\IEEEauthorrefmark{3}pawan.kumar@iiit.ac.in}}
\maketitle

\maketitle

\begin{abstract}
A parallel and nested version of a frequency filtering preconditioner is proposed for linear systems corresponding to diffusion equation on a structured grid. The proposed preconditioner is found to be robust with respect to jumps in the diffusion coefficients. The storage requirement for the preconditioner is $O(N),$ where $N$ is number of rows of matrix, hence, a fairly large problem of size more than 42 million unknowns has been solved on a quad core machine with 64GB RAM. The parallelism is achieved using twisted factorization and SIMD operations. The preconditioner achieves a speedup of 3.3 times on a quad core processor clocked at 4.2 GHz, and compared to a well known algebraic multigrid method, it is significantly faster in both setup and solve times for diffusion equations with jumps.
\end{abstract}


\section{Introduction}
\label{sec:introduction}
We consider the problem of solving large sparse linear systems of the form 
\begin{align}
    Ax = b, \quad A \in \mathbb{R}^{m \times m}, \quad b \in \mathbb{R}^m, \label{eqn:linsys}
\end{align}
which arises, for example, during the numerical solution of the following diffusion equation
\begin{equation}
\begin{aligned}
    -\text{div} (\kappa(x) \nabla u) &= f \quad \text{in}~\Omega, \\
    u &= 0 \quad \text{on}~\partial \Omega_D, \label{eqn:diffeqn} \\
    \dfrac{\partial u}{\partial n} &= 0 \quad \text{on}~\partial \Omega_N.
\end{aligned}
\end{equation}
Here $\Omega$ is the interior of the domain, and $\partial \Omega_D$ and $\partial \Omega_N$ are the Dirichlet and Neumann boundaries respectively. 
The matrix $A$ in \eqref{eqn:linsys} is assumed to be symmetric and positive definite, and the ideal solver is the preconditioned conjugate gradient method~\cite{saad1996} (or PCG in short), where the convergence of the method depends on the quality of the preconditioner, or more precisely on the condition number of $B^{-1}A,$ where $B$ is the preconditioner. For large jumps in the diffusion coefficient $\kappa(x),$ the gradient methods are slow to converge, unless an efficient preconditioner is used. Moreover, with the modern day hardware consisting of multicore and manycore architectures, it is desirable to have a solver capable of exploiting parallelism available in the hardware to converge rapidly to a desired solution. In this paper, we propose a fast multithreaded preconditioned gradient solver for the diffusion problem on a structured grid.

Given that the problem \eqref{eqn:diffeqn} arises in a wide variety of scientific simulations, it has been studied widely. For these problems, there are three main classes of methods that have been studied: domain decomposition methods \cite{toselli,kumar2012}, multigrid methods \cite{kumar2015,amg,kumar2014a,kumar2013}, block preconditioners \cite{kumar2016a}, deflation preconditioners \cite{kumar2020, kumar2021, kumar2016b}. The method proposed in this paper belongs to this third class, with some ideas of multigrid. Speaking of block preconditioners, they may be further classified into those that are specifically designed for structured grids, and those that are designed without considering any structure in the computational grid. 

On a structured grid, the discretization schemes such as finite difference, and finite volume methods lead to a ``nested" block tridiagonal matrix. Exploiting this structure is essential to obtaining a scalable and memory efficient solver. One of the fastest solvers known today for solving \eqref{eqn:diffeqn} are Algebraic Multigrid Methods (AMG in short)~\cite{amg,tro,hypre}. The AMG solvers exploit matrix structure extremely well, and scale to several thousands of cores \cite{baker2012}.

The filtering preconditioners are constructed such that they can filter out some undesirable components from the error during iterative solve~\cite{Nat,Ach,buz1998,buz2004,wag1997a,wag1997b}. A preconditioner $B$ is said to satisfy filtering property when 
\begin{align}
    Bt = At, \label{eqn:filter}
\end{align}
where $t$ is a given filter vector. When 
\begin{align}
    t = {\bf 1} = [1,1,\dots, 1]^T, \label{rowsum}
\end{align}
then \eqref{eqn:filter} is called the rowsum constraint, i.e., the preconditioner $B$ is constructed such that the sum of a row of $B$ is equal to the sum of the corresponding row of $A.$ A classical preconditioner based on rowsum constraint is modified incomplete ILU \cite{saad1996}. The preconditioner proposed in this paper tries to satisfy a similar constraint as \eqref{eqn:filter}. To motivate filtering preconditioner, we recall that the expression for the error for a fixed point iteration at the $(n+1)$th step denoted by $e^{n+1}$ is given as follows: 
\begin{align*}
    e^{n+1} &= (I - B^{-1}A)e^n \\ 
    &= (I - B^{-1}A)^2 e^{n-1} \\ 
    &= \cdots \cdots \cdots \cdots \\
    &= (I - B^{-1}A)^{n+1}e^0. 
\end{align*}
Clearly, if $B$ satisfies the filter condition \eqref{eqn:filter}, then $$(I - B^{-1}A)t=0,$$ and by choosing a suitable filter vector $t,$ a desired component of the error vector is removed; we may express the error vector as a linear combination of the eigenvectors of $A$ (assuming $A$ is SPD). The components of the error corresponding to the eigenvectors corresponding to small eigenvalues of $B^{-1}A$ are most difficult to damp out from error. Hence, if the filter vectors are chosen to be some approximation to the eigenvector corresponding to smallest eigenvalues, then the filtering preconditioner will help in faster convergence of the preconditioned conjugate gradient method. For unstructured grid, a multithreaded implementation was previously investigated in \cite{kumar2012,kumar2014}, and a distributed memory variant was studied in \cite{Qu2013}. To the best of our knowledge, none of the implementations of filtering preconditioners, or similar deflation preconditioners shown so far could compete with state-of-the-art AMG solvers. In this paper, we propose a new nested filtering preconditioner, and show that a parallel implementation can significantly outperform AMG on quad cores. The parallelism is achieved using twisted factorization and SIMD operations. Since we only store bands of the preconditioner, the memory requirement for the preconditioner is $O(N),$ where $N$ is number of rows of matrix. More importantly, our solver does not involve any parameters, hence, we have ignored comparisons with geometric multigrid, or other solvers that require a lot of parameter tuning. We show that our implementation is in most cases significantly faster than the popular AMG solver \cite{hypre} in both setup and solve time. 

\section{The Nested and Twisted Frequency Filtering Preconditioner}
For a 3D structured $nx \times ny \times nz$ grid where, $nx$ denotes the number of points on the line, $ny$ denotes the number of lines on each plane, and $nz$ denotes the number of planes we have a $(nx \times ny \times nz) \times (nx \times ny \times nz)$ finite difference matrix as shown below. Let $nxy$ denote $nx \times ny,$ the number of unknowns on each plane, $nyz$ denote total number of lines on all planes 
and $nxyz$ denote $nx \times ny \times nz,$ the total number of unknowns. 
\begin{align*}
 A = \left(\begin{array}{llll}
\widehat{D}_{1} & \widehat{U}_{3}^{1} &  &   \\ 
\widehat{L}_{3}^{1} & \widehat{D}_{2} & \ddots &   \\ 
 & \ddots & \ddots& \widehat{U}_{3}^{nz-1}  \\ 
 &  & \widehat{L}_{3}^{nz-1} & \widehat{D}_{nz}  \\ 
 \end{array}\right). 
\end{align*}
Here the diagonal blocks $\widehat{D}_{i}'s$ are the blocks corresponding 
to the unknowns in the $i^{th}$ plane, and the blocks $\widehat{L}_{3}^{i}$
and $\widehat{U}_{3}^{i}$ are diagonal matrices of size $nxy$ and they 
correspond to the connections between the $i^{th}$ and $(i+1)^{th}$ plane.
We assume that the diagonal blocks $\widehat{D}_{i}$ are themseleves 
block tridiagonal, i.e., the blocks $\widehat{D}_{i}$ are denoted by,
\begin{align*}
 \widehat{D}_{i} = \left(\begin{array}{llll}
\overline{D}_{(i-1)*ny+1} & \overline{U}_{2}^{(i-1)*ny+1} &  &   \\ 
\overline{L}_{2}^{(i-1)*ny+1} & \overline{D}_{(i-1)*ny+2} & \ddots &   \\ 
 & \ddots & \ddots & \overline{U}_{2}^{(i*ny-1}  \\ 
 &  & \overline{L}_{2}^{i*ny-1} & \overline{D}_{i*ny}  \\ 
 \end{array}\right), 
\end{align*} the blocks $\overline{L}_{2}^{j}$ and 
$\overline{U}_{2}^{j}$ 
are diagonal matrices of size $nx,$ and they correspond to the 
connections between 
the line blocks $\overline{D}_{j}$ and $\overline{D}_{j+1}$. We further assume that the diagonal blocks, 
$\overline{D}_{i}$ are 
themseleves tridiagonal matrices
\begin{align*}
\overline{D}_{i} = \left(\begin{array}{llll}
\tilde{D}_{(i-1)*nx+1} & \tilde{U}_{1}^{(i-1)*nx +1} &  &  \\ 
\tilde{L}_{1}^{(i-1)*nx+1} & \tilde{D}_{(i-1)*nx+2} & \ddots &  \\ 
 & \ddots & \ddots & \tilde{U}_{i*nx-1} \\ 
 &  & \tilde{L}_{1}^{i*nx-1} &\tilde{D}_{i*nx}
\end{array}\right)
\end{align*}
with the scalars $\tilde{L}_{1}^{i*nx + j}$ and 
$\tilde{U}_{1}^{i*nx + j}$ 
being the corresponding connections between the cells of the line.

\subsection{Construction of Twisted Filtering Preconditioner}
To expose parallelism in the Nested Low Frequency Tangential Filtering Decomposition, 
we twist the matrices $U$ and $L$ about the main diagonal, flipping one half of the subdiagonal (or superdiagonal) to the other side. As we will see shortly, each such twist gives us a two-way parallelism, and it is possible to perform a twist at each level (or dimension) in the nested hierarchy. This gives us a 8-way parallelism approach in theory. The structure of the matrices $L_{1}$ and $U_{1}$ after twisting is shown below.
\begin{align*}
L_{1}  &= \left(\begin{array}{llllllll} 
 0 & 0 & & & & & \\
\tilde{L}^{1}_{1} & \ddots & \ddots & & & \\ 
& \ddots & \ddots & 0 & & \\
& & \tilde{L}^{mid}_1 & \ddots & \tilde{L}^{mid+1}_1 &\\
& & & 0 & \ddots & \ddots \\
& & & & \ddots & \ddots & \tilde{L}^{nxyz-1}_1 \\
& & & & & 0 & 0 \\
\end{array} \right), \\
U_{1}  &= \left(\begin{array}{llllllll} 
 0 & \tilde{U}^{1}_{1} & & & & & \\
 0 & \ddots & \ddots & & & \\ 
& \ddots & \ddots & \tilde{U}^{mid}_1 & & \\
& & 0 & \ddots & 0 &\\
& & & \tilde{U}^{mid+1}_1 & \ddots & \ddots \\
& & & & \ddots & \ddots & 0 \\
& & & & & \tilde{U}^{nxyz-1}_1 & 0 \\
 \end{array} \right)
 \end{align*}
 

The other twisted matrices $L_2, U_2, L_3,$ and $U_3$ are defined similarly. 
\begin{align*} 
L_{2}  &= \left(\begin{array}{llllllll} 
 0 & 0 & & & & & \\
\overline{L}^{1}_{2} & \ddots & \ddots & & & \\ 
& \ddots & \ddots & 0 & & \\
& & \overline{L}^{mid} & \ddots & \overline{L}^{mid+1}_2 &\\
& & & 0 & \ddots & \ddots \\
& & & & \ddots & \ddots & \overline{L}^{nyz-1}_2 \\
& & & & & 0 & 0 \\
 \end{array} \right), \\
U_{2}  &= \left(\begin{array}{llllllll} 
 0 & \overline{U}^{1}_{2} & & & & & \\
 0 & \ddots & \ddots & & & \\ 
& \ddots & \ddots & \overline{U}^{mid}_2 & & \\
& & 0 & \ddots & 0 &\\
& & & \overline{U}^{mid+1}_2 & \ddots & \ddots \\
& & & & \ddots & \ddots & 0 \\
& & & & & \overline{U}^{nyz-1}_2 & 0 \\
 \end{array} \right),
 \end{align*}
 and the twisted 3rd level lower and upper bands are given by 
 \begin{align*}
L_3  &= \left(\begin{array}{llllllll} 
 0 & 0 & & & & & \\
\widehat{L}^{1}_3 & \ddots & \ddots & & & \\ 
& \ddots & \ddots & 0 & & \\
& & \tilde{L}^{mid}_3 & \ddots & \widehat{L}^{mid+1}_3 &\\
& & & 0 & \ddots & \ddots \\
& & & & \ddots & \ddots & \widehat{L}^{nz-1}_3 \\
& & & & & 0 & 0 \\
 \end{array} \right), \\
U_3  &= \left(\begin{array}{llllllll} 
 0 & \widehat{U}^{1}_{3} & & & & & \\
 0 & \ddots & \ddots & & & \\ 
& \ddots & \ddots & \widehat{U}^{mid}_3 & & \\
& & 0 & \ddots & 0 &\\
& & & \widehat{U}^{mid+1}_3 & \ddots & \ddots \\
& & & & \ddots & \ddots & 0 \\
& & & & & \widehat{U}^{nz-1}_3 & 0 \\
 \end{array} \right). 
\end{align*}

In actual implementation, we only need to store the bands of A.


To create the preconditioner, we first consider the block LU factorization 
$$A = (P + L_{3})(I + P^{-1}U_{3}).$$

The $A$ in this equation is already known to us, and on simplifying the right hand side, and solving for diagonal blocks $P_i$ of $P,$ we get the following recurrence solution for $P_{i}$ 
\begin{equation}
\begin{aligned}
P_{i} = \begin{cases}
\widehat{D}_{1}, \quad i~ =~ 1,\\
\widehat{D}_{i} - \widehat{L}_{3}^{i-1}(P_{i-1}^{-1})\widehat{U}_{3}^{i-1}, \quad i = 2, \cdots, j-1, \\
\widehat{D}_{nz}, \quad i~ =~ nz,\\
\widehat{D}_{i} - \widehat{L}_{3}^{i}(P_{i+1}^{-1})\widehat{U}_{3}^{i}, \quad i = nz-1, \cdots, j+1, \\
\widehat{D}_{i} - \widehat{L}_{3}^{i-1}(P_{i-1}^{-1})\widehat{U}_{3}^{i-1} - \widehat{L}_{3}^{i}(P_{i+1}^{-1})\widehat{U}_{3}^{i}, \quad i = j.  
\end{cases}
\end{aligned}
\end{equation}

In the above iteration, as $i$ increases, $P_i$ tends to become denser, hence, it is costly to compute terms such as $\widehat{L}_{3}^{i-1}(P_{i-1}^{-1})\widehat{U}_{3}^{i-1}.$ Moreover, storing $P_i$ is costly, hence, we will replace $P_i^{-1}$ by its sparse approximation. Reusing the notation $P_i$ for approximated $P_i,$ we define the following approximation to $P_i$
\begin{equation}
\begin{aligned}
P_{i} = \begin{cases}
\widehat{D}_{1}, \quad i~ =~ 1,\\
\widehat{D}_{i} - \widehat{L}_{3}^{i-1}(2\beta_{i-1}-\beta_{i-1}{P}_{i-1}
\beta_{i-1})\widehat{U}_{3}^{i-1}, \\ \quad i = 2, \cdots, j-1, \\
\widehat{D}_{nz}, \quad i~ =~ nz,\\
\widehat{D}_{i} - \widehat{L}_{3}^{i}(2\beta_{i+1}-\beta_{i+1}{P}_{i+1}
\beta_{i+1})\widehat{U}_{3}^{i}, \\ \quad i = nz-1, \cdots, j+1, \\
\widehat{D}_{i} - \widehat{L}_{3}^{i-1}(2\beta_{i-1}-\beta_{i-1}{P}_{i-1}
\beta_{i-1})\widehat{U}_{3}^{i-1} \\ \hspace{0.5cm} - \widehat{L}_{3}^{i}(2\beta_{i+1}-\beta_{i+1}{P}_{i+1}
 \beta_{i+1})\widehat{U}_{3}^{i}, \quad i = j.  
\end{cases}
\end{aligned}
\end{equation}
Here $j$ is the block row index where the twist happens, $\beta_{i}$ are diagonal matrices defined as
\begin{align*}
\beta_i = \text{diag}((P_{i-1}^{-1}\hat{U}^{i-1}) ./ (\hat{U}^{i-1}\hat{t}_i)),
\end{align*}
where $\hat{t}_i$ is a vector of all ones,
and $$2\beta_{i-1}-\beta_{i-1}P_{i-1}\beta_{i-1},$$ or $$2\beta_{i+1}-\beta_{i+1}P_{i+1}\beta_{i+1}$$ for the lower half is claimed to be a better approximation to $({P_{i}})^{-1}$. Note that the product on the rhs no longer equals $A$ after substituting $P^{-1}$ with it's $\beta$ approximated form. After approximation, we define the NTD preconditiner $B_{\text{NTD}}$ as follows
\begin{align}
B_{\text{NTD}} = (P + L_{3})(I + P^{-1}U_{3}). \label{eqn:precon}
\end{align}

Since $\beta_i's$ are diagonals, the sparsity pattern of $P_i$ is same as that of $\widehat{D}_i.$ Hence, like $\hat{D_i}$ blocks, the individual $P_i$ blocks are themselves nested block tridiagonal, we can obtain a further factorization as follows 
\begin{align}
P = (T + L_{2})(I + {T}^{-1}U_{2}). 
\end{align}

As for $P_i$ blocks before, we have the following recurrence solution for $T_i$ blocks
\begin{align}
T_{i} = \begin{cases}
 \overline{D}_{1}, \quad i~ =~ 1,\\
\overline{D}_{i} - \overline{L}_{2}^{i-1}(2\beta_{i-1}-\beta_{i-1}{T}_{i-1}
\beta_{i-1})\overline{U}_{2}^{i-1}, \\ \quad i = 2, \cdots, j-1, \\
\overline{D}_{nz}, \quad i~ =~ nz, \\
\overline{D}_{i} - \overline{L}_{2}^{i}(2\beta_{i+1}-\beta_{i+1}{T}_{i+1}
\beta_{i+1})\overline{U}_{2}^{i}, \\ \quad  i = nz-1, \cdots, j+1, \\
\overline{D}_{i} {-} \overline{L}_{2}^{i-1}(2\beta_{i-1}-\beta_{i-1}{T}_{i-1}
\beta_{i-1})\overline{U}_{2}^{i-1} \\ \hspace{0.5cm} {-} \overline{L}_{2}^{i}(2\beta_{i+1}-\beta_{i+1}{T}_{i+1}
\beta_{i+1})\overline{U}_{2}^{i},  \quad i = j,
\end{cases}
\end{align}
where $j$ is the block row index, $\beta_{i}'s$ are diagonal matrices defined as 
\begin{align*}
\beta_i = \text{diag}((T_{i-1}^{-1}\overline{U}^{i-1}) ./ (\overline{U}^{i-1}\bar{t}_i)),
\end{align*}
where $\bar{t}_i$ is vector of all ones,
and as shown above, we consider the approximation $$2\beta_{i-1}-\beta_{i-1}T_{i-1}\beta_{i-1}$$ for upper half, and similarly the approximation $$2\beta_{i+1}-\beta_{i+1}T_{i+1}\beta_{i+1}$$ for the lower half is claimed to be a better approximation to $({T_{i}})^{-1}.$ 

Again sparsity pattern of $T_i$ is same as $\bar{D}_i,$ i.e., the $T_i$ blocks are themselves pointwise tridiagonal matrices, and can be approximated similarly as follows: 
\begin{align}
T = (M + L_{1})(I + M^{-1}U_{1}). 
\end{align}

Since $T$ is block diagonal with tridiagonal blocks, the above factorization is exact. We obtain the recurrence for $M_i$ as follows:
\begin{align}
M_{i} =
\begin{cases}
 \tilde{D}_{1}, \quad i~ =~ 1,\\
\tilde{D}_{i} - \tilde{L}_{1}^{i-1}{M}_{i-1}^{-1}\tilde{U}_{1}^{i-1},~ i = 2, \cdots, j-1, \\
\tilde{D}_{nz}, \quad i~ =~ nz,\\
\tilde{D}_{i} - \tilde{L}_{1}^{i}{M}_{i+1}^{-1}\tilde{U}_{1}^{i}, \quad i = nz-1, \cdots, j+1, \\
\tilde{D}_{i} {-} \tilde{L}_{1}^{i-1}{M}_{i-1}^{-1}\tilde{U}_{1}^{i-1} - \tilde{L}_{1}^{i}{M}_{i+1}^{-1}\tilde{U}_{1}^{i}, \quad i = j, 
\end{cases}
\end{align}
where $j$ is the row index, and $M_{i-1}^{-1}$ (or $M_{i+1}^{-1}$) is reciprocal of $M_{i-1}$ (or $M_{i+1}$).
Note that during construction of the preconditioner, we only need to store the bands as follows:
\begin{align*}
 \ell_3 &= [\text{ThirdNonzeroLowerBand}(B), 0 \dots, 0], \\
 u_3 &= [\text{ThirdNonzeroUpperBand}(B), 0 \dots, 0],\\
 \ell_2 &= [\text{SecondNonzeroLowerBand}(P), 0 \dots, 0], \\
 u_2 &= [\text{SecondNonzeroUpperBand}(P), 0 \dots, 0],\\
 \ell_1 &= [\text{FirstNonzeroLowerBand}(T), 0], \\
 u_1 &= [\text{FirstNonzeroUpperBand}(T), 0].
\end{align*}
Here $$\ell_1, \ell_2, \ell_3, u_1, u_2, u_3$$ are vectors of length $N,$ with appropriate zero padding at the end. Note that to extract these bands, we do not construct the matrices $T, P,$ and $B.$ We extract these bands during the recurrence for $T_i$ and $P_i.$ Also, the outermost bands $\ell_3$ and $u_3$ are same as the outermost bands of $A.$

\subsection{Solve Routine}
\begin{algorithm*}
\caption{NTD Solve}\label{eqn:ntd_solve}
\begin{algorithmic}[1]
\Function{Solve}{$start$, $end$, $x$, $b$, $block\_size$, $level$}
\If {$level = 1$} // base case \\
	\hspace{1cm} Perform iterative solve using SIMD operations, see section \eqref{sec:simd} 
\EndIf
\State LowerSolve($start$, $end$, $x$, $b$, $block\_size$, $level$);
\State UpperSolve($start$, $end$, $x$, $b$, $block\_size$, $level$); 
\EndFunction
\end{algorithmic}
\end{algorithm*}

\begin{algorithm*}
\caption{Lower Solve}\label{LowerSolve}
\begin{algorithmic}[1]
\Function{LowerSolve}{$start$, $end$, $x$, $b$, $block\_size$, $level$}
\State $num\_blocks = (end{-}start{+}1)/block\_size$
\State $cur = start{+}\floor{(num\_blocks-1)/2}*block\_size$ // $mid\_block\_start$
\State $prev = cur {-} block\_size$
\State $next = cur {+} block\_size$ 
\State LowerSolveUpperHalf($start$, $cur{-}1$, $x$, $b$, $block\_size$, $level$) \label{call_to_lowersolveupperhalf}
\State LowerSolveLowerHalf($next$, $end$, $x$, $b$, $block\_size$, $level$) \label{call_to_lowersolvelowerhalf}
\State $L \gets \ell_{level}$
\State $b(cur{:}next{-}1) = b(cur{:}next{-}1) {-} L(prev{:}cur{-}1) .* x(prev{:}cur{-}1)$
\State $b(cur{:}next{-}1) = b(cur{:}next{-}1) {-} L(cur{:}next{-}1) .* x(next{:}next{+}block\_size{-}1)$
\State \text{SOLVE}($cur$, $next{-}1$, $x$, $b$, $block\_size/n$, $level{-}1$)
\EndFunction
\end{algorithmic}
\end{algorithm*}

\begin{algorithm*}
\caption{Lower Solve Upper Half}\label{LowerSolveUpperHalf}
\begin{algorithmic}[1]
\Function{LowerSolveUpperHalf}{$start$, $end$, $x$, $b$, $block\_size$, $level$}
\State \text{SOLVE}($start$, $start{+}block\_size{-}1$, $x$, $b$, $block\_size/n$, $level{-}1$, $n$)
\For{$i$ in $start{+}block\_size{:}end{:}block\_size$}
\State $L \gets \ell_{level}$
\State $j \gets i\text{{+}block\_size{-}1}$
\State $b(i{:}j) = b(i{:}j) {-} L(i{-}block\_size{:}i{-}1).*x(i{-}block\_size{:}i{-}1)$
\State \text{SOLVE}($i$, $j$, $x$, $b$, $block\_size/n$, $level{-}1$)
\EndFor
\EndFunction
\end{algorithmic}
\end{algorithm*}

\begin{algorithm*}
\caption{Lower Solve Lower Half}\label{LowerSolveLowerHalf}
\begin{algorithmic}[1]
\Function{LowerSolveLowerHalf}{$start$, $end$, $x$, $b$, $block\_size$, $level$}
\State $idx \gets \textit{end{-}block\_size{+}1}$
\State \text{SOLVE}($idx$, $end$, $x$, $b$, $block\_size/n$, $level-1$)
\For{$i$ in $end{-}2*block\_size{+}1{:}start{:}{-}block\_size$}
\State $L \gets \ell_{level}$
\State $j \gets \textit{i{+}block\_size{-}1}$
\State $b(i{:}j) = b(i{:}j) {-} L(i{:}i{+}block\_size{-}1).*x(i{:}i{+}block\_size{-}1)$
\State \text{SOLVE}($i$, $j$, $x$, $b$, $block\_size/n$, $level{-}1$)
\EndFor
\EndFunction
\end{algorithmic}
\end{algorithm*}

\begin{algorithm*}
\caption{Upper Solve}\label{UpperSolve}
\begin{algorithmic}[1]
\Function{UpperSolve}{$start$, $end$, $x$, $b$, $block\_size$, $level$}
\State $num\_blocks = (end{-}start{+}1)/block\_size$
\State $cur = start{+}\floor{(num\_blocks-1)/2}*block\_size$ // $mid\_block\_start$
\State $next = cur {+} block\_size$
\State $x(cur{:}next{-}1) = b(cur{:}next{-}1)$
\State UpperSolveUpperHalf($start$, $cur{-}1$, $x$, $b$, $block\_size$, $level$)
\State UpperSolveLowerHalf($next$, $end$, $x$, $b$, $block\_size$, $level$)
\EndFunction
\end{algorithmic}
\end{algorithm*}

\begin{algorithm*}
\caption{Upper Solve Upper Half}\label{UpperrSolveUpperHalf}
\begin{algorithmic}[1]
\Function{UpperrSolveUpperHalf}{$start$, $end$, $x$, $b$, $block\_size$, $level$}
\State $x' \gets zeros(1{:}N)$
\State $b' \gets zeros(1{:}N)$
\For{$i$ in $end{-}block\_size{+}1{:}start{:}{-}block\_size$}
\State $U \gets u_{level}$
\State $j \gets \textit{i{+}block\_size{-}1}$
\State $b'(i{:}j) = U(i{:}j) \, .* \, x'(j{+}1{:}j{+}block\_size)$
\State \text{SOLVE}($i$, $j$, $x'$, $b'$, $block\_size/n$, $level{-}1$)
\State $x(i{:}j) = b(i{:}j) {-} x'(i{:}j)$
\EndFor
\EndFunction
\end{algorithmic}
\end{algorithm*}

For simplicity, let us assume that $nx=ny=nz=n,$ and we define $N = n^3$, denoting the number of rows of the matrix. To solve $Bx = y$ using block LU form in \eqref{eqn:precon}, we use the Algorithm \ref{eqn:ntd_solve} by calling the {\tt SOLVE} routine in it as ${\tt SOLVE}(1, N, x, y, n^2, 3)$. Here, in the arguments to the {\tt SOLVE} function, $start$ and $end$ denote the start and end position of the slice of solution vector $x$ being solved for against slice of rhs vector $b$. The three recursive levels of the solve are identified by the argument $level$, which reduces by 1, each time we recurse into Solve, while the $block\_size$ denotes the size of block at the current level. It is assumed that the functions are able to access the preconditioner bands and the value $n$ from the global scope. Subsequently, in the functions {\tt LOWERSOLVE} and {\tt UPPERSOLVE}, the solve for upper and lower halves can be invoked in parallel due to the parallelism exposed by twisting performed.
\begin{algorithm*}
\caption{Upper Solve Lower Half}\label{UpperSolveLowerHalf}
\begin{algorithmic}[1]
\Function{UpperSolveLowerHalf}{$start$, $end$, $x$, $b$, $block\_size$, $level$}
\State $x' \gets zeros(1{:}N)$ 
\State $b' \gets zeros(1{:}N)$
\For{$i$ in $start{:}end{:}block\_size$} 
\State $U \gets u_{level}$
\State $j \gets \textit{i{+}block\_size{-}1}$
\State $b'(i{:}j) = U(i{-}block\_size{:}i{-}1)\, {.*} \, x(i{-}block\_size{:}i{-}1)$
\State \text{SOLVE}($i$, $j$, $x'$, $b'$, $block\_size/n$, $level{-}1$)
\State $x(i{:}j) = b(i{:}j) {-} x'(i{:}j)$
\EndFor
\EndFunction
\end{algorithmic}
\end{algorithm*}

Let us look at function {\tt LOWERSOLVE} in Algorithm \ref{LowerSolve} in detail, and function {\tt UPPERSOLVE} in Aglorithm \ref{UpperSolve} can be understood similarly. In the function {\tt LOWERSOLVE}, $cur$ refers to the row index of first element of block at which twist is performed, and similarly $prev$ and $next$ denote the row index of first element in the block previous and next to this block. Due to independence of two halves, we invoke their respective solves in step \ref{call_to_lowersolveupperhalf} and step \ref{call_to_lowersolvelowerhalf} in parallel. The synchronization is done in step 9, 10, and 11 by computing the solution corresponding to the block at which twist has been performed. Inside the calls to {\tt LOWERSOLVEUPPERHALF} and {\tt LOWERSOLVELOWERHALF}, we iterate from top to middle, and middle to bottom respectively, and call the {\tt SOLVE} routine in Algorithm \ref{eqn:ntd_solve} recursively by reducing $level$ to $level{-}1$ and $block\_size$ to $block\_size/n$. This recursion ends at the base case in step 2 in Algorithm 1 {\tt SOLVE} routine which is achieved when we reach level 1, in which case we perform the solve with pointwise tridiagonal matrix by leveraging SIMD vector operations instead of going further down into the recursion. It is worth noting that although the independence between the two halves holds at each level down the hierarchy of nested block tridiagonal structure and in theory we can split into two parallel calls at each level, but as we go deeper into the hierarchy, the problem size keeps reducing and at the innermost level the time taken to do the actual work present is so less that it becomes comparable to the time required to create two threads and later synchronize them at the merge step. Hence, instead of doing the twist at innermost level and invoking two threads, we do a SIMD Solve leveraging the vectorized operations to realize more speedup as has been described in more detail in the subsequent section. If implemented in a naive way, at each $level$ in recursion, we would have to create and destroy two threads as many times as a matrix is solved at that level, which equals the number of blocks in the main diagonal for that level. This number translates to $n$ for the second level (the last level till which we create threads), which linearly grows with the problem size. To avoid such an overhead, we observe that throughout the lifetime of the algorithm atmost 4 threads are active at a time. Hence instead of using the recursive algorithm as described above, we invoke 4 threads at the entry point of the solver, which coordinate among themselves using conditional variables to execute the recursive algorithm in a bottom up fashion such that we incur minimum ovehead due to threading and extract maximum speedup.

\subsection{\label{sec:simd}A Note on SIMD Operations in Innermost Solve} 
For the innermost level, let us consider the case of {\tt LOWERSOLVE} when the matrix only has the main diagonal and subdiagonal (the diagonal below main diagonal) and similar explanation would apply for innermost {\tt UPPERSOLVE}. Let's denote this matrix as $T$, the solution vector as $x$, and right hand side vector as $b$. Let the value at an arbitrary position $i$ (1-indexed notation) in $x$ be denoted by $x[i]$. Note that for $i>1,$ $x[i]$ can be written in terms of $x[i-1],$ because
\begin{align*}
T[i][i-1]*x[i-1] + T[i][i]*x[i] = b[i].
\end{align*}
For the first entry, the following equation holds true:
\begin{align*}
T[1][1] * x[1] = b[1].
\end{align*}
Using induction, we can compute $x[i]$, where $i > 1$, but we need to compute $x[i-1]$ before we can compute $x[i]$.
CPUs have SIMD instruction sets (like SSE and AVX). AVX allows performing 256 bit sized operations. This allows fitting four 64-bit "double" values. For vectorization, note that without actually knowing $x[i-1]$, we can find constants $a$ and $b$ such that $$x[i] = a*x[i-1] + b.$$ We store $a$ and $b$ instead, and do the task of actually finding $x[i]$ later.
\begin{align*} 
	x[i+1] &= a'*x[i] + b' \\
		   &= a'*(a*x[i-1] + b) + b' \\
		   &= a'*a*x[i-1] + a'b + b' \\
		   &= c*x[i-1] + d.
\end{align*}
Note that we have expressed $x[i+1]$ in terms of $x[i-1]$ by having 2 coefficients, $c$ and $d$.
This can be further extended for $x[i+2]$ and so on.
Thus, to vectorize, we assume $K=4$ equally spaced elements in $x$, and find next $N/K$ values in terms of the assumed values, where $N$ is equal to size of $A$ ($A$ is assumed to be a square matrix).
Expressing our equations in these terms allows us to write $x[i]$ in every block of size $N/K$ in form of $c*y+z$ where $c$ is a constant, and x, y, and z can be treated as vectors by the CPU to optimize this operation using AVX instructions like Fused-Multiply Add. We'd have to do more compute operations here, first while finding the coefficients, and then while substituting the coefficients to find $x$. Also note that since we're finding only in terms of assumed $K$ values, to compute the final $x,$ we'd need $N/K$ vector operations of size $K$.

%
%
%

\subsubsection{Memory Requirement:} 
We recall that the size of given matrix $A$ is $N \times N,$ where $N=nxyx,$ and since we operate on sparse diagonal storage format, we need $\mathcal{O}(N)$ memory.

\section{Numerical Experiments}
All the results shown have been obtained by running the experiments on intel i7-7700K CPU with 4 physical cores, 64 GB DDR4 RAM. The compiler version used is {\tt gcc 7.3} with {\tt -march=native} and {\tt -O3} flags. To measure time, we have used {\tt tick\_count} class from {\tt intel TBB} 2018 update 4. For BLAS operations, we have used {\tt intel MKL} 2018 update 3. The Hypre version used is 2.8 along with {\tt metis-5.1.0} for graph partitioning, and OpenMPI library. 
\begin{center}
\begin{table*}[t]
\caption{\label{tab:NTD} Time (in seconds) and Iterations for NTD+ILU0 preconditioner}
\centering
\begin{tabular}{ |p{1.2cm}||p{1.3cm}|p{1.8cm}|p{1.5cm}|p{1.5cm}|p{1.5cm}|p{0.8cm}|}
 \hline
 \multicolumn{7}{|c|}{\bf{NTD}} \\
 \hline
\bf{Matrix Type} & \bf{Number of Rows} & \bf{Relative Residual} & \bf{Setup Time} & \bf{Solve Time} & \bf{Overall Time} & \bf{Iters} \\
 \hline
1 & 1000000 & 1.00E-07 & 1.67 &  0.75 & 2.42 & 16 \\
1 & 8000000 & 1.00E-07 & 13.53 & 8.64 &  22.18 & 22 \\
1 & 42875000 & 1.00E-07 & 82.80 & 52.90 &  135.70 & 26 \\
2 & 1000000 & 1.00E-07 & 1.69 & 0.76 & 2.45 & 16 \\
2 & 8000000 & 1.00E-07 & 13.43 & 8.03 &  21.47 & 21 \\
2 & 42875000 & 1.00E-07 & 82.55 & 50.72 &  133.28 & 25 \\
3 & 1000000 & 1.00E-07 & 1.66 & 0.76 & 2.43 & 16 \\
3 & 8000000 & 1.00E-07 & 13.50 & 8.10 &  21.60 & 21 \\
3 & 42875000 & 1.00E-07 & 82.78 & 50.86 &  133.64 & 25 \\
1 & 1000000 & 1.00E-10 & 1.70 & 1.34 &  3.05 &  30 \\
1 & 8000000 & 1.00E-10 & 13.38 & 16.42 &  29.80 &  44 \\
1 & 42875000 & 1.00E-10 & 82.74 & 108.06 &   190.80 & 55 \\
2 & 1000000 & 1.00E-10 & 1.68 & 1.18 &  2.86 &  26 \\
2 & 8000000 & 1.00E-10 & 13.40 & 13.14 &   26.54 &  35 \\
2 & 42875000 & 1.00E-10 & 83.01 & 85.06 &  168.07 & 43 \\
3 & 1000000 & 1.00E-10 & 1.68 & 1.18 &  2.86 & 26 \\
3 & 8000000 & 1.00E-10 & 13.56 & 13.17 &  26.73 &  35 \\
3 & 42875000 & 1.00E-10 & 82.82 & 85.21 &  168.03 & 43 \\
 \hline
\end{tabular}
\end{table*}
\end{center}

\subsubsection{Test Matrices:} We consider the boundary value problem \eqref{eqn:diffeqn} on a unit cube $$\Omega=[0,1] \times [0,1] \times [0,1]$$ with zero Dirichlet boundary condition for all the test problems. We consider the uniform grid, i.e. $nx=ny=nz.$ We choose $nx = 100, 200,$ and $350.$ To test the preconditioner, we consider the following three test cases. 
\begin{enumerate}
\item {\bf Skyscrapper problems with checkerboard variable jumps:} Here the domain contains many zones of high permeability which are isolated from each other. Let $[x]$ denote the integer value of $x.$ Here the tensor $\kappa(x)$ is defined as follows:


\begin{align*}
\kappa(x) = \begin{cases}
10^3 \times ([10 \times x_2] + 1), \\
\quad \text{if}~[10x_i] \equiv 0~(\text{mod}\, 2), \quad i=1,2,3, \\ \\
1, \quad \text{otherwise}.
\end{cases}
\end{align*}

We call this the Type 1 matrix.
\item {\bf Non-homogeneous problems with jump in coefficients:} Here the tensor $\kappa$
is isotropic and discontinuous. It jumps from the constant value $10^3$ in the ring $$1/2\sqrt 2 \leq |x-c| \leq 1/2, \, c = (1/2, 1/2, 1/2)^T$$ to $1$ outside. We call this the Type 2 matrix.
\item {\bf Poisson Problem:} Here the tensor $\kappa=1.$ We call this the Type 3 matrix.
\end{enumerate}

\subsubsection{Smoothing high frequency components of error:} As in multigrid, we use incomplete LU factorization of given matrix $A$ with no fill-in as a smoother and combine it with NTD preconditioner described previously. The combination preconditioner denoted by $B_{\text{c}}$ can be defined as follows:
\begin{align}
B_c^{-1} = B_{\text{NTD}}^{-1} + B_{\text{ILU0}}^{-1} - B_{\text{NTD}}^{-1} A B_{\text{ILU0}}^{-1}, \label{eqn:com}
\end{align}
where $B_{\text{ILU0}}$ denotes the ILU preconditioner \cite{saad1996}. Note that such a combination was proposed and explored in \cite{Nat,kumar2015}.

\subsubsection{Parallel Solve with Preconditioner:} From the equation \eqref{eqn:com} above, we notice that solving with the preconditioner $B_c$ requires solving with $B_{\text{NTD}}$ and $B_{\text{ILU0}},$ and a matrix vector multiplication with the given coefficient matrix $A.$ We have showed how to solve with $B_{\text{NTD}}$ on a quad core in previous sections. The solve with $B_{\text{ILU0}}$ requires triangular solves, i.e. forward sweep followed by a backward sweep, which are inherently sequential in nature. To address this bottleneck, instead of doing ILU0 for full matrix $A,$ we do an ILU0 of a block diagonal approximation $\tilde{A}$ of matrix A, where 


\[\tilde{A} = \text{blkD} (A (1:M, 1:M ), A (M+1:N, M+1:N )),\]
where $M = \text{floor}(N/2),$ and blkD stands for block diagonal.

Such an approximation does not lead to any degradation in the performance of the combination preconditioner, and allowed us to engage two threads during the construction and solve phase of $B_\text{ILU0}.$  For the matrix times vector operation, we have implemented a parallel banded matrix vector multiply routine engaging 4 threads, further leveraging \text{SIMD} operations inside each thread. 
\begin{center}
\begin{table*}[t]
\caption{\label{tab:Hypre Parameters} Parameters used for Hypre}
\centering
\begin{tabular}{ |p{1.2cm}||p{1.4cm}|p{1.4cm}|p{1.8cm}|p{1.6cm}|p{1.4cm}|p{1.4cm}|}
 \hline
\bf{Number of Rows} & \bf{Coarsen Type} & \bf{Relax Type} & \bf{Strong Threshold} & \bf{Agg Num Levels} & \bf{Interp Type} & \bf{Trunc Factor} \\
\hline
1000000 & 10 & 6 & 0.25 & 8 & 3 & 0.1 \\
8000000 & 6 & 5 & 0.25 & 0 & 0 & 0 \\
42875000 & 3 & 6 & 0.5 & 0 & 0 & 0 \\
\hline
\end{tabular}
\end{table*}
\end{center}
\subsubsection{Parameters for conjugate gradient:} All the experiments have been done using preconditioned conjugate gradient method with $B_c$ as a preconditioner and we compare it Hypre (a V-cycle of Boomer AMG) preconditioned CG. The maximum number of iterations was set to 200, and the iterations were stopped when the relative residual denoted by $$\| Ax^{i} - b \|_2/ \| b \|_2$$ became less than $tol$, where we considered the following values for $tol$: $10^{-7}, 10^{-10}.$ Here, $x^i$ denotes the approximate solution at the $i$th iteration of PCG, and the initial solution $x^0$ is a vector of all zeros. 

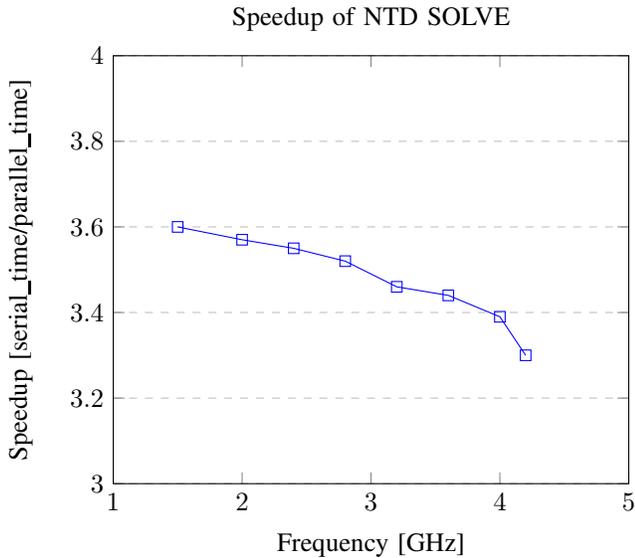
\begin{figure}
\begin{center}
\begin{tikzpicture}
\begin{axis}[
    title={Speedup of NTD SOLVE},
    xlabel={Frequency [GHz]},
    ylabel={Speedup [serial\_time/parallel\_time]},
    xmin=1, xmax=5,
    ymin=3, ymax=4,
    legend pos=north west,
    ymajorgrids=true,
    grid style=dashed,
]
 
\addplot[
    color=blue,
    mark=square,
    ]
    coordinates {
    (1.5, 3.6) (2.0, 3.57) (2.4, 3.55) (2.8, 3.52) (3.2, 3.46) (3.6, 3.44) (4.0, 3.39) (4.2, 3.3)
    };
\end{axis}
\end{tikzpicture}
\end{center}
\caption{\label{fig:speedup}Speedup on quad core versus cpu clock frequency.}
\end{figure}

Table \ref{tab:NTD} and Table \ref{tab:Hypre} show the results using {\tt NTD + ILU0} preconditioner and {\tt HYPRE} respectively. The exact parameters used for Hypre have been shown in Table \ref{tab:Hypre Parameters}, which have been fine tuned for the solve time by performing rigorous and expensive search experiments in the parameter space, and we believe they give the best possible solve time (not necessarily the best setup time) with Hypre with 4 MPI processes for the respective matrix sizes. The function NTD Solve, which has been parallelized using 4 threads, shows a maximum speedup of 3.6x. The amount of speedup realized reduces with increasing cpu frequency because the routine involves many operations such as elementwise vector multiply, elementwise vector add among others, which are inherently memory bound. In such algorithms, we do not benefit from increased amount of CPU compute power available on increasing the number of parallel cores since the rate at which memory supplies the data becomes the bottleneck. However, the trend observed in Figure \ref{fig:speedup} shows that theoretically the proposed algorithm  is strong enough to realize speedup with faster memory chips. On comparing the two methods, we find that in most cases the proposed method is faster than Hypre in both solve and setup times.
\begin{center}
\begin{table*}[t]
\caption{\label{tab:Hypre} Time (in seconds) and Iterations for Hypre preconditioner}
\centering
\begin{tabular}{ |p{1.2cm}||p{1.3cm}|p{1.8cm}|p{1.5cm}|p{1.5cm}|p{1.5cm}|p{0.8cm}|}
 \hline
 \multicolumn{7}{|c|}{\bf{Hypre}} \\
 \hline
\bf{Matrix Type} & \bf{Number of rows} & \bf{Relative Residual} & \bf{Setup Time} & \bf{Solve Time} & \bf{Overall Time} & \bf{Iters} \\
 \hline
1 & 1000000 & 1.00E-07 & 3.07 & 0.94 & 4.01 &  12 \\
1 & 8000000 & 1.00E-07 & 202.04 & 14.89 &  216.94 &  9 \\
1 & 42875000 & 1.00E-07 & 1842.56 & 369.86 &  2212.42 &  42 \\
2 & 1000000 & 1.00E-07 & 3.15 &  0.78 & 3.93 &  10 \\
2 & 8000000 & 1.00E-07 & 206.84 &  9.89 &  216.73 & 6 \\
2 & 42875000 & 1.00E-07 & 1802.17 & 68.25 &  1870.42 & 8 \\
3 & 1000000 & 1.00E-07 & 3.16 & 0.78 & 3.95 &  10 \\
3 & 8000000 & 1.00E-07 & 207.04 &  9.95 &  216.99 & 6 \\
3 & 42875000 & 1.00E-07 & 1800.41 & 68.30 &  1868.71 & 8 \\
1 & 1000000 & 1.00E-10 & 3.07 & 1.37 &   4.45 &  18 \\
1 & 8000000 & 1.00E-10 & 202.45 & 20.84 &  223.30 &  13 \\
1 & 42875000 & 1.00E-10 & 1842.14 & 525.02 &  2367.16 &  60 \\
2 & 1000000 & 1.00E-10 & 3.14 & 1.07 &  4.21 &  14 \\
2 & 8000000 & 1.00E-10 & 206.94 & 14.15 &  221.09 &  9 \\
2 & 42875000 & 1.00E-10 & 1804.47 & 91.15 & 1895.62 &  11 \\
3 & 1000000 & 1.00E-10 & 3.15 & 1.07 &  4.23 &  14 \\
3 & 8000000 & 1.00E-10 & 207.15 & 14.23 &  221.38 &  9 \\
3 & 42875000 & 1.00E-10 & 1810.21 & 91.03 &  1901.24 & 11 \\
 \hline
\end{tabular}
\end{table*}
\end{center}


\begin{thebibliography}{00}

\bibitem{App} Appleyard, J.R., Cheshire, I.M.: Nested Factorization, SPE, (1983)
 
\bibitem{Nat} Achdou, Y., Nataf, F.: Low Frequency Tangential Filtering Decomposition, \textbf{14}(2), Num. Lin. Alg. App., (2006)
 
\bibitem{Ach} Achdou, Y., Nataf, F.: Dimension wise Iterated Frequency Filtering Decomposition, Numerical Linear Algebra with Applications, (2001)

\bibitem{Axel} Axelson, O., Polman, O.: A robust preconditioner based on algebraic substructuring and two level grids, In: Hackbusch, W.(ed.) Robust multi-grid methods, NNFM, Bd.23. Vieweg-Verlag, Braunschweig.

\bibitem{Gabr} Wittum, G.: Filternde Zerlegungen- Schnelle Loeser fur grosse Gleichungssysteme, Teubner Skripten zur Numerik Band 1, Teubner-Verlag, Stuttgart (1992)

\bibitem{buz1998} Buzdin, A.: Tangential decomposition. Computing, \textbf{61}(3), 257--276, (1998)

\bibitem{buz2004} Buzdin, A., Wittum, G.: Two-frequency decomposition. Numer. Math., \textbf{97}(2), 269--295 (2004)

\bibitem{wag1997a} Wagner, C.:. Tangential frequency filtering decompositions for symmetric matrices, Numer. Math., \textbf{78}(1), 119--142 (1997)

\bibitem{wag1997b} Wagner, C.: Tangential frequency filtering decompositions for unsymmetric matrices. Numer. Math.. 78(1), 143--163 (1997)

\bibitem{saad1996} Saad, Y.: Iterative Methods for Sparse Linear Systems. PWS publishing company, Boston, MA, 1996.

\bibitem{amg} Stuben, K.: A review of algebraic multigrid, J. Comput. Appl. Math. and applied
mathematics, \textbf{128}(1-2), 281--309, (2001)

\bibitem{tro} Trottenberg, U., Osterlee, C. W., Schuller, A.: Multigrid, Academic Press, (2000)

\bibitem{hypre} HYPRE: High Performance Preconditioners, \url{https://computation.llnl.gov/projects/hypre-scalable-linear-solvers-multigrid-methods}

\bibitem{kumar2010} Kumar, P.: A Class of Parallel Preconditioning Techniques Suitable for Partial Differential Equations Defined on Structured and Unstructured Mesh, PhD thesis, \url{http://www.theses.fr/2010PA112118}, (2010)

\bibitem{kumar2012} Kumar P., Meerbergen K., Roose D.: Multi-threaded Nested Filtering Factorization Preconditioner. In: Manninen P., Öster P. (eds) Applied Parallel and Scientific Computing. PARA 2012. Lecture Notes in Computer Science, vol 7782. Springer, Berlin, Heidelberg

\bibitem{kumar2015} Kumar, P., Grigori, L., Niu, Q., Nataf, F.: On relaxed nested factorization and combination preconditioning, International Journal of Computer Mathematics, (2015)

\bibitem{toselli} Toselli, A., Widlund, O. B.: Domain Decomposition Methods - Algorithms and Theory, Springer, Berlin, Heidelberg, \url{https://doi.org/10.1007/b137}

\bibitem{kumar2014} Kumar, P., Multithreaded Direction Preserving Preconditioners, ISPDC, 2014

\bibitem{Qu2013} Qu, L., Grigori, L., Nataf, F., Parallel Design and Performance of Nested Filtering Factorization Preconditioner, SC, 2013

\bibitem{baker2012} Baker A.H., Falgout R.D., Kolev T.V., Yang U.M., Scaling Hypre’s Multigrid Solvers to 100,000 Cores. In: Berry M. et al. (eds) High-Performance Scientific Computing. Springer, London (2012)

\bibitem{kumar2021} Shrutimoy Das, Siddhant Katyan, Pawan Kumar; Proceedings of the IEEE/CVF Winter Conference on Applications of Computer Vision (WACV), 2021, pp. 1782-1789

\bibitem{kumar2020} Siddhant Katyan, Shrutimoy Das, Pawan Kumar; Proceedings of the IEEE/CVF Winter Conference on Applications of Computer Vision (WACV), 2020, pp. 3588-3595

\bibitem{kumar2016a} Kumar P. (2016) Fast Preconditioned Solver for Truncated Saddle Point Problem in Nonsmooth Cahn–Hilliard Model. In: Fidanova S. (eds) Recent Advances in Computational Optimization. Studies in Computational Intelligence, vol 655. Springer, Cham. https://doi.org/10.1007/978-3-319-40132-4\_10

\bibitem{kumar2014a} Pawan Kumar (2014) Aggregation based on graph matching and inexact coarse grid solve for algebraic two grid, International Journal of Computer Mathematics, 91:5, 1061-1081, DOI: 10.1080/00207160.2013.821115


\bibitem{niu2010} Niu, Q., Grigori, L., Kumar, P. et al. Modified tangential frequency filtering decomposition and its fourier analysis. Numer. Math. 116, 123–148 (2010). https://doi.org/10.1007/s00211-010-0298-3

\bibitem{kumar2016b} Pawan kumar, Preconditioners based on approximation of non-standard norms for phase separation applications, AIP Conference Proceedings 1738, 480114 (2016)

\bibitem{kumar2013} Pawan Kumar, Stefano Markidis, Giovanni Lapenta, Karl Meerbergen, Dirk Roose,
High Performance Solvers for Implicit Particle in Cell Simulation,
Procedia Computer Science, Volume 18, 2013, Pages 2251-2258.

\end{thebibliography}
\end{document}